\documentclass[12pt]{article}
\usepackage{epic}
\def\option{\noindent}
\def\vs{\vspace}
\def\Tr{{\rm Tr}}
\def\mod{{\rm \;mod\;}}
\def\gcd{{\rm  gcd}}
\def\lcm{{\rm  lcm}}
\def\e{{\rm e}}
\def\ie{{\it i.e.}}

\newcommand{\Z}{\mathsf{Z}\kern -5pt \mathsf{Z}}
\def\half{ {1\over 2} }
\def\twelfth{ {1\over 12} }
\def\id{ 0 } 
\def\power{m}
\def\subpower{l}
\def\a{\alpha}
\def\b{\beta}
\def\Dr{{ \Delta }}
\def\Lam{\Lambda}
\def\lam{\lambda}
\def\sig{\sigma}
\def\om{\omega}
\def\omc{{\omega_c}}
\def\Zcl{ Z^{\rm closed} }
\def\Zop{ Z^{\rm open} }
\def\xNK{ {x_{N,K}} }
\def\xNN{ {x_{N,N}} }
\def\xKN{ {x_{K,N}} }
\def\xnk{ {x_{2n+1,2k+1}} }
\def\xnn{ {x_{2n+1,2n+1}} }
\def\xkn{ {x_{2k+1,2n+1}} }
\def\Pplus{ P_+^K }
\def\zeroth{zero${}^{\rm th}$ }

\def\foreven{{\rm \;for\;} N+K {\rm \; even~(except~}N=K=2^\power)}
\def\except{{\rm \;for\;} N+K {\rm \; even~(except~for~}N=K=2^\power)}
\def\el{{\ell_1}}
\def\pirho{{\pi(\rho)}}
\def\cH{  {\cal H}  }
\def\Exp{  {\cal E}^{\om}  }
\def\Bound{  {\cal B}^{\om}  }
\def\Boundomc{  {\cal B}^{\omc}  }
\def\taa{{\tilde{a}}}
\def\ta{{\tilde{\a}}}
\def\tb{{\tilde{\b}}}
\def\tell{{\tilde{\ell}}}
\def\tel{{\tilde{\el}}}
\def\tn{{\tilde{n}}}
\def\tN{{\tilde{N}}}
\def\tq{{\tilde{q}}}
\def\tQ{{\tilde{Q}}}
\def\tS{{\tilde{S}}}
\def\tchi{{\tilde{\chi}}}
\def\tlam{{\tilde{\lambda}}}
\def\tLam{{\tilde{\Lambda}}}
\def\tmu{{\tilde{\mu}}}
\def\tnu{{\tilde{\nu}}}
\def\tpsi{{\tilde{\psi}}}
\def\trho{{\tilde{\rho}}}
\def\rhohat{{\hat{\rho}}}
\def\pirhohat{{\pi(\rhohat)}}
\def\hS{{S'}}
\def\htS{{\tS'}}
\def\htSstar{{\tS^{\prime *}}}
\def\hlam{{\hat{\lambda}}}
\def\bz{{\bar z}}

\def\bV{\overline{V}}
\def\bJ{\overline{J}}
\def\bL{\overline{L}}
\def\bI{\overline{I}}
\def\bJ{\overline{J}}
\def\g{g}
\def\gcup{ {\breve g} }
\def\suN{{{\rm su}(N)}}
\def\sun{{{\rm su}(n+1)}}
\def\suK{{{\rm su}(K)}}
\def\supm{{{\rm su}(p^\power)}}
\def\suoddn{{{\rm su}(2n+1)}}
\def\suoddk{{{\rm su}(2k+1)}}
\def\spn{{{\rm sp}(n)}}
\def\spk{{{\rm sp}(k)}}
\def\hgK{{\hat{g}_K}}   
\def\twhgK{{\hat{g}_K^\om}}   
\def\twhgk{{\hat{g}_{2k+1}^\omc}}   
\def\suNK{{\widehat{\rm su}(N)_K}}
\def\soNK{{\widehat{\rm so}(N)_K}}
\def\sunK{{\widehat{\rm su}(n+1)_K}}

\def\suoddnk{{\widehat{\rm su}(2n+1)_{2k+1}}}
\def\suoddkn{{\widehat{\rm su}(2k+1)_{2n+1}}}
\def\suNN{{\widehat{\rm su}(N)_N}}
\def\sutt{{\widehat{\rm su}(3)_3}}
\def\suKN{{\widehat{\rm su}(K)_N}}
\def\soKN{{\widehat{\rm so}(K)_N}}
\def\AnoneK{  (A_{n}^{(1)})_K}
\def\Annonek{  (A_{2n}^{(1)})_{2k+1}}
\def\Anntwok{  (A_{2n}^{(2)})_{2k+1}}
\def\spnK{{\widehat{\rm sp}(n)_K}}
\def\spnk{{\widehat{\rm sp}(n)_k}}
\def\spkn{{\widehat{\rm sp}(k)_n}}
\def\ConeK{  (C_n^{(1)})_K}
\def\Conek{  (C_n^{(1)})_k}
\def\bdy{  | B \rangle\!\rangle }
\def\twbdy{  | B \rangle\!\rangle^\om }

\def\ishimu{  | \mu \rangle\!\rangle_I }
\def\twishimu{  | \mu \rangle\!\rangle_I^\om }
\def\Ctwishimu{  | \mu \rangle\!\rangle_I^\omc }
\def\ishinu{  | \nu \rangle\!\rangle_I }
\def\twishinu{  | \nu \rangle\!\rangle_I^\om }
\def\ishimubra{  {}_I\langle \! \langle  \mu | } 
\def\twishimubra{  {}_I^\om\langle \! \langle  \mu | } 
\def\cardymu{  | \mu \rangle\!\rangle_C }
\def\cardya{  | \a \rangle\!\rangle_C^\om }
\def\Ccardya{  | \a \rangle\!\rangle_C^\omc }
\def\cardyb{  | \b \rangle\!\rangle_C^\om }
\def\cardylam{  | \lam \rangle\!\rangle_C }
\def\cardylambra{  {}_C \langle \! \langle  \lam  | } 
\def\cardyabra{  {}_C^\om \langle \! \langle  \a  | } 

\def\Nabc{ {N_{\mu\nu}}^{\lam}  }
\def\tNabcD{ {{\tN}_{\tmu \tnu}}^{~~\sig^\Dr(\tlam) } }
\def\tNabc{ {{\tN}_{\tmu \tnu}}^{~~\tlam } }
\def\nabc{ {n_{\mu\nu}}^{\lam}  }
\def\nblama{ {n_{\b\lam}}^{\a}  }
\def\four{  {\vcenter  {\vbox  
              {\hrule height.4pt
               \hbox {\vrule width.4pt  height3pt  
                      \kern3pt 
                      \vrule width.4pt  height3pt 
                      \kern3pt
                      \vrule width.4pt  height3pt 
                      \kern3pt
                      \vrule width.4pt  height3pt 
                      \kern3pt
                      \vrule width.4pt height3pt}
               \hrule height.4pt}
                         }
              }
           }
\def\oneoneoneone{ 
              {\vcenter  {\vbox  
              {\hrule height.4pt
               \hbox {\vrule width.4pt  height3pt  
                      \kern3pt 
                      \vrule width.4pt  height3pt }
               \hrule height.4pt
               \hbox {\vrule width.4pt  height3pt  
                      \kern3pt 
                      \vrule width.4pt  height3pt }
               \hrule height.4pt
               \hbox {\vrule width.4pt  height3pt  
                      \kern3pt 
                      \vrule width.4pt  height3pt }
               \hrule height.4pt
               \hbox {\vrule width.4pt  height3pt  
                      \kern3pt 
                      \vrule width.4pt  height3pt }
               \hrule height.4pt}
                         }
              }
           }
\def\be{\begin{equation}}
\def\ee{\end{equation}}
\def\bea{\begin{eqnarray}}
\def\eea{\end{eqnarray}}

\def\theequation{\thesection.\arabic{equation}}

\begin{document}
\bibliographystyle{bst}

\begin{flushright}
{\tt hep-th/0601175}\\
BRX-TH-572\\
BOW-PH-136\\
\end{flushright}
\vspace{30mm}

\vspace*{.3in}

\begin{center}
{\Large\bf\sf  Level-rank duality of
untwisted and twisted D-branes}

\vskip 5mm Stephen G. Naculich\footnote{Research supported in part
by the NSF under grant PHY-0456944}$^{,a}$
and Howard J.  Schnitzer\footnote{Research supported in part 
by the DOE under grant DE--FG02--92ER40706\\
{\tt \phantom{aaa} schnitzr@brandeis.edu; naculich@bowdoin.edu}\\
}$^{,b}$

\end{center}

\begin{center}
$^{a}${\em Department of Physics\\
Bowdoin College, Brunswick, ME 04011}

\vspace{.2in}

$^{b}${\em Martin Fisher School of Physics\\
Brandeis University, Waltham, MA 02454}
\end{center}
\vskip 2mm

\begin{abstract}
Level-rank duality of untwisted and twisted D-branes 
of WZW models is explored.
We derive the relation between D0-brane charges of 
level-rank dual untwisted D-branes of $\suNK$ and $\spnk$,
and of level-rank dual twisted D-branes of $\suoddnk$.
The analysis of level-rank duality of twisted D-branes of $\suoddnk$
is facilitated by their close relation to untwisted D-branes of $\spnk$.
We also demonstrate level-rank duality of the 
spectrum of an open string stretched between untwisted 
or twisted D-branes in each of these cases.
\end{abstract}

\vfil\break


\section{Introduction}
\renewcommand{\theequation}{1.\arabic{equation}}
\setcounter{equation}{0}

D-branes on group manifolds have been the subject of much work,
both from the algebraic and geometric points of view
\cite{Klimcik:1996hp}--\cite{Gaberdiel:2004yn}.
(For a review, see ref.~\cite{Schomerus:2002dc}.
See also ref.~\cite{Halpern:2004jm}.)
Algebraically, these D-branes correspond to 
the allowed boundary conditions for a Wess-Zumino-Witten (WZW)
model on a surface with boundary \cite{Cardy:1989ir}.

Much can be learned about D-branes by studying their charges,
which are classified by K-theory or,
in the presence of a cohomologically nontrivial $H$-field background,
twisted K-theory 
\cite{Minasian:1997mm}.
The charge group for D-branes 
on a simply-connected group manifold $G$
with level $K$ is given by the twisted K-group
\cite{Fredenhagen:2000ei, 
Maldacena:2001xj,
Bouwknegt:2000qt,
Freed:2001jd, 
Braun:2003rd,
Gaberdiel:2004hs}
\be
\label{eq:Ktheory}
K^*(G) = \oplus_{i=1}^{m}
 \Z_x\,,\qquad
m= 2^{{\rm rank}\,G - 1}
\ee
where $\Z_x \equiv \Z/x\Z$ 
with $x$ an integer depending on $G$ and $K$. 
For $\suNK$, for example, $x$ is given 
by \cite{Fredenhagen:2000ei}
\be
\label{eq:xnk}
\xNK \equiv {  N+K \over \gcd \{ N+K, \lcm \{ 1, \ldots, N-1\} \}  }\,.
\ee
One of the $\Z_x$ factors in the charge group corresponds 
to the charge of untwisted (symmetry-preserving) D-branes.
For $\suN$ with $N>2$, another of the $\Z_x$ factors corresponds to 
D-branes twisted by the charge-conjugation symmetry.
For the D-branes corresponding to the remaining factors, see 
refs.~\cite{Fredenhagen:2000ei,Maldacena:2001xj,Gaberdiel:2004hs}.

WZW models with classical Lie groups 
possess an interesting property called level-rank duality: 
a relationship between various quantities in the 
$\suNK$, $\soNK$, or $\spnk$ model, 
and corresponding quantities in the level-rank dual 
$\suKN$, $\soKN$, or $\spkn$ model
\cite{Naculich:1990hg}--\cite{Bourdeau:1991uu}.
Implications of level-rank duality
for boundary Kazama-Suzuki models
were explored in ref.~\cite{Ishikawa:2003kk}.

In ref.~\cite{Naculich:2005tn},
we began the study of level-rank duality in boundary WZW theories, 
and in particular the level-rank duality of untwisted D-branes of $\suNK$.
In this paper, we extend this work 
to untwisted D-branes of the  $\spnk$ WZW model, 
and to twisted D-branes of $\suoddnk$,
which are closely related to the untwisted D-branes of $\spnk$.
We focus on two aspects of this duality:
the relation between the D0-brane charges of 
level-rank dual D-branes,
and the level-rank duality of the 
spectrum of an open string stretched between 
untwisted or twisted D-branes
(\ie, the coefficients of the open-string partition function). 
For untwisted D-branes, these coefficients are given
by the fusion coefficients of the bulk WZW theory \cite{Cardy:1989ir},
so duality of the untwisted open-string partition function
follows from the well-known level-rank duality of the fusion 
rules \cite{Naculich:1990hg,Altschuler:1989nm,Mlawer:1990uv}.
For twisted D-branes, the open-string partition function coefficients
may be calculated in terms of the  
modular-transformation matrices of twisted affine Lie 
algebras \cite{Birke:1999ik,Ishikawa:2001zu,Gaberdiel:2002qa}.
In this paper, we show that the
spectrum of an open string stretched between 
twisted D-branes of $\suoddnk$ is level-rank dual.

In section 2, we review some 
salient features of untwisted D-branes of WZW models. 
Section 3 describes the level-rank duality of the charges 
of untwisted D-branes of $\suNK$ for all values of $N$ and $K$  
(our results in ref.~\cite{Naculich:2005tn} were restricted to $N+K$ odd),
and of the untwisted open-string partition function.
Section 4 describes the level-rank duality 
of the charges of untwisted D-branes of $\spnk$,
and of the untwisted open-string partition function.
Twisted D-branes of WZW models are reviewed in section 5,
and section 6 is devoted to demonstrating the level-rank duality 
of the charges of twisted D-branes of $\suoddnk$,
and of the twisted open-string partition function.
Concluding remarks constitute section 7.

\section{Untwisted D-branes of WZW models} 
\renewcommand{\theequation}{2.\arabic{equation}}
\setcounter{equation}{0}

In this section, we review some salient features 
of Wess-Zumino-Witten models
and their untwisted D-branes.

The WZW model,
which describes strings propagating on a group manifold,
is a rational conformal field theory
whose chiral algebra (for both left- and right-movers)
is the (untwisted) affine Lie algebra $\hgK$ at level $K$. 
The Dynkin diagram of $\hgK$ has one more node 
than that of the associated finite-dimensional Lie algebra $\g$.
Let $(m_0, m_1, \cdots, m_n)$ be the dual Coxeter labels of $\hgK$ 
(where $n = {\rm rank~} \g$)
and $h^\vee = \sum_{i=0}^n m_i$ the dual Coxeter number of $\g$.
The Virasoro central charge  of the WZW model is then
$c ={K \dim \g}/(K+h^\vee)$.

The building blocks of the WZW conformal field theory
are integrable highest-weight representations 
$V_\lam$ of $\hgK$, 
that is, representations whose highest weight $\lam \in \Pplus$ 
has non-negative Dynkin indices $(a_0, a_1, \cdots, a_n)$ 
satisfying
\be
\label{eq:integrable}
\sum_{i=0}^n  m_i a_i = K\,.
\ee
With a slight abuse of notation, we also use $\lam$ to denote 
the highest weight of the irreducible representation of $\g$ 
with Dynkin indices $(a_1, \cdots, a_n)$,
which spans the lowest-conformal-weight subspace of $V_\lam$.

For $\sunK = \AnoneK$ and $\spnK = \ConeK$, 
the untwisted affine Lie algebras with which 
we will be principally concerned, we have
$m_i=1$ for $i=0, \cdots, n$, and $h^\vee = n+1$.
It is often useful to describe irreducible representations 
of $\g$ in terms of Young tableaux.
For example,
an irreducible representation of $\sun$ or $\spn$
whose highest weight $\lam$ has Dynkin indices $a_i$ 
corresponds to a Young tableau with $n$ or fewer rows, 
with row lengths 
\be
\ell_i = \sum_{j=i}^{n} a_j \,, \qquad 
i=1, \ldots, n  \,.
\ee
Let $r(\lam) = \sum_{i=1}^{n} \ell_i$ denote the number of boxes
of the tableau.
Representations $\lam$ 
corresponding to integrable highest-weight representations
$V_\lam$ of $\sunK$ or ~$\spnK$
have Young tableaux with $K$ or fewer columns.

We will only consider WZW theories with a diagonal closed-string spectrum:
\be
\label{eq:diagonal}
\cH^{\rm closed}  
= \bigoplus_{\lam \in \Pplus}  
 V_\lam \otimes \bV_{\lam^*}
\ee
where $\bV$ represents right-moving states,
and $\lam^*$ denotes the representation conjugate to $\lam$.
The partition function for this theory is
\be
\label{eq:closedpartition}
\Zcl (\tau) 
= \sum_{\lam \in \Pplus} \left|  \chi_\lam (\tau)  \right|^2
\ee
where
\be
\chi_\lam (\tau) = \Tr_{V_\lam} q^{L_0 - c/24}\,, \qquad 
q =  \e^{2\pi i \tau}
\ee
is the affine character of the integrable highest-weight 
representation $V_\lam$.
The affine characters transform linearly
under the modular transformation $\tau \to -1/\tau$,
\be 
\label{eq:modulartrans}
\chi_\lam(-1/\tau) = \sum_{\mu \in \Pplus} S_{\mu\lam} \; \chi_\mu(\tau)\,,
\ee
and the unitarity of $S$ ensures
the modular invariance of the partition function 
(\ref{eq:closedpartition}).

Next we turn to consider D-branes in the WZW model
\cite{Klimcik:1996hp}-\cite{Gaberdiel:2004yn}.
These D-branes may be studied algebraically 
in terms of the possible boundary conditions 
that can consistently be imposed on a WZW model with boundary.
We consider boundary conditions that leave unbroken the $\hgK$ symmetry, 
as well as the conformal symmetry, of the theory,
and we label the allowed boundary conditions 
(and therefore the D-branes) by $\a$, $\b$, $\cdots$.
The partition function on a cylinder,
with boundary conditions $\a$ and $\b$ on the two boundary components, 
is then given as a linear combination of 
affine characters of $\hgK$ \cite{Cardy:1989ir}
\be
\label{eq:bdypartition}
\Zop_{\a\b} (\tau) = \sum_{\lam \in \Pplus} \nblama \chi_\lam (\tau) \,.
\ee
This describes the spectrum of
an open string stretched between D-branes labelled by $\a$ and $\b$. 

In this section, 
we consider a special class of boundary conditions,
called {\it untwisted} (or {\it symmetry-preserving}),
that result from imposing the restriction
\be
\label{eq:untwistedconditions}
\left[ J^a(z) - \bJ^a(\bz)\right] \bigg|_{z=\bz} = 0
\ee
on the currents of the affine Lie algebra on the boundary $z=\bz$
of the open string world-sheet, 
which has been conformally transformed to the upper half plane.
Open-closed string duality allows one to correlate 
the boundary conditions (\ref{eq:untwistedconditions}) 
of the boundary WZW model 
with coherent states $\bdy \in \cH^{\rm closed}$ 
of the bulk WZW model satisfying
\be
\label{eq:modes}
\left[  J^a_m + \bJ^a_{-m} \right] \bdy = 0\,, \qquad m\in \Z
\ee
where $J^a_m$ are the modes of the affine Lie algebra generators.
Solutions of eq.~(\ref{eq:modes})
that belong to a single sector $V_\mu \otimes \bV_{\mu^*}$
of the bulk WZW theory 
are known as Ishibashi states $\ishimu$ \cite{Ishibashi:1988kg},
and are normalized such that 
\be
\label{eq:ishinorm}
\ishimubra q^H \ishinu =  \delta_{\mu\nu} \chi_\mu (\tau)\,, 
\qquad q = \e^{2\pi i \tau}
\ee
where $H = \half \left( L_0 + \bL_0 - \twelfth c \right)$
is the closed-string Hamiltonian.
For the diagonal theory (\ref{eq:diagonal}),
Ishibashi states exist for all integrable highest-weight 
representations $\mu \in \Pplus$ of $\hgK$.

A coherent state $\bdy$ that corresponds to an
allowed boundary condition
must also satisfy additional (Cardy) conditions \cite{Cardy:1989ir},
among which are that the coefficients $\nblama$ in 
eq.~(\ref{eq:bdypartition}) must be non-negative integers.
Solutions to these conditions 
are labelled by integrable highest-weight representations 
$\lam \in \Pplus$ of the untwisted affine Lie algebra $\hgK$,
and are known as (untwisted) Cardy states $\cardylam$.
The Cardy states may be expressed as linear combinations 
of Ishibashi states 
\be
\label{eq:cardyishi}
\cardylam = \sum_{\mu \in \Pplus}  
{S_{\lam \mu} \over \sqrt{S_{\id\mu}}} \ishimu 
\ee
where $S_{\lam\mu}$ is the modular transformation matrix 
given by eq.~(\ref{eq:modulartrans}),
and $\id$ denotes the identity representation.
Untwisted D-branes of $\hgK$ correspond to $\cardylam$ 
and are therefore also labelled by $\lam \in \Pplus$.

The partition function of open strings stretched between 
untwisted D-branes $\lam$ and $\mu$ 
\be
\label{eq:openpart}
\Zop_{\lam\mu} (\tau) = \sum_{\nu \in \Pplus} \nabc \chi_\nu (\tau) 
\ee
may alternatively be calculated as the closed-string propagator between
untwisted Cardy states \cite{Cardy:1989ir}
\be
\label{eq:prop}
\Zop_{\lam\mu} (\tau) = \cardylambra \tq^H \cardymu\,, 
\qquad \tq = \e^{2\pi i (-1/\tau)   }\,.
\ee
Combining eqs.~(\ref{eq:prop}), 
(\ref{eq:cardyishi}),
(\ref{eq:ishinorm}), 
(\ref{eq:modulartrans}),
and the Verlinde formula \cite{Verlinde:1988sn},
we find 
\be
\label{eq:verlinde}
\Zop_{\lam\mu} (\tau) 
= \sum_{\rho \in \Pplus}
{ S^*_{\lam\rho} S_{\mu\rho} \over S_{\id\rho} } \chi_\rho(-1/\tau)
= \sum_{\nu \in \Pplus} \sum_{\rho \in \Pplus}
{ S_{\mu\rho} S_{\nu\rho} S^*_{\lam\rho} \over S_{\id\rho} }  \chi_\nu(\tau)
= \sum_{\nu \in \Pplus} \Nabc \chi_\nu(\tau)\,.
\ee
Hence, the coefficients $\nabc$ in the 
open-string partition function  (\ref{eq:openpart})
are simply given by the fusion coefficients $\Nabc$ of the bulk WZW model. 

Finally, an untwisted D-brane labelled by $\lam \in \Pplus$
can be considered a bound state of D0-branes \cite{Affleck:1990by, 
Alekseev:1999bs, 
Bachas:2000ik, 
Alekseev:2000jx,
Fredenhagen:2000ei, 
Maldacena:2001xj}.
It possesses a conserved D0-brane charge $Q_\lam$ given by $(\dim \lam)_\g$,
but the charge is only defined modulo some integer \cite{Alekseev:2000jx,
Fredenhagen:2000ei,
Maldacena:2001xj,
Bouwknegt:2002bq}.
For D-branes of $\suNK$, for example, this integer is given
by eq.~(\ref{eq:xnk}), thus 
\be
 Q_\lam = (\dim \lam)_\suN \quad \mod \xNK \qquad {\rm for}~~ \suNK
\ee
is the charge of the untwisted D-brane labelled by $\lam$.

\section{Level-rank duality of untwisted D-branes of $\suNK$} 
\renewcommand{\theequation}{3.\arabic{equation}}
\setcounter{equation}{0}

In ref.~\cite{Naculich:2005tn}, 
the relation between the charges of untwisted D-branes of the $\suNK$ model 
and those of the level-rank-dual $\suKN$ model was ascertained
for odd values of $N+K$.
In this section, we extend these results to all values of $N$ and $K$.

Since charges of $\suNK$ D-branes are only defined modulo $\xNK$,
and those of $\suKN$ D-branes modulo $\xKN$,
comparison of charges of level-rank-dual D-branes 
is only possible modulo $\gcd \{ \xNK, \xKN  \} $.
Without loss of generality we will henceforth 
assume that $N \ge K$,
in which case 
$\gcd \{ \xNK, \xKN  \} = \xNK $.

\vs{.1in}
\noindent{\bf Level-rank duality of untwisted D-brane charges}
\vs{.1in}

\option
Given a Young tableau $\lam$ corresponding to an integrable highest-weight
representation of $\suNK$ (with $N-1$ or fewer rows, and $K$ or fewer columns),
its transpose $\tlam$ corresponds to an integrable highest-weight
representation of $\suKN$.
(The map between representations of $\suNK$ and $\suKN$ is not one-to-one,
but the map between cominimal equivalence classes of representations is.
These equivalence classes are generated by the simple-current symmetry 
$\sigma$ of $\suNK$, which takes $\lam$ into $ \lam' = \sig (\lam)$,
where the Dynkin indices of $\lam'$ are 
$ a'_i = a_{i-1} $ for $ i=1, \ldots, N-1,$ and $ a'_0 = a_{N-1}$.)

For odd $N+K$, the relation between $Q_\lam$, 
the charge of the untwisted $\suNK$ D-brane labelled by $\lam$,
and $\tQ_\tlam$, 
the charge of the level-rank-dual $\suKN$ D-brane labelled by $\tlam$,
was shown to be \cite{Naculich:2005tn} 
\be
\label{eq:oddduality}
\tQ_\tlam = (-1)^{r(\lam)}  Q_\lam \quad \mod \xNK,  
\qquad {\rm for~} N+K {\rm ~odd}\,.  \qquad\qquad\qquad\qquad
\ee
where $r(\lam)$ is the number of boxes in the tableau $\lam$.
In this section,
we show that for the case of even $N+K$, the charges obey
\be
\label{eq:evenduality}
\tQ_\tlam = Q_\lam \quad \mod  \xNK, 
\qquad \except. 
\ee
In the remaining case,
we conjecture the relation 
\be
\label{eq:specialduality}
\tQ_\tlam = \left\{ 
{ (-1)^{r(\lam)/N} Q_\lam  \quad  \mod \xNN,~~~ {\rm~when~}  N~|~r(\lam) \atop
                   Q_\lam  \quad  \mod \xNN,~~~ {\rm~when~}  N~\not |~r(\lam) }
\right\} \quad
{\rm~for~}  N=K=2^\power 
\ee
for which we have numerical evidence, but (as of yet) no complete proof. 

\vs{.1in}
\noindent {\it Proof of eq.~(\ref{eq:evenduality}):}   
We proceed as in ref.~\cite{Naculich:2005tn} 
by writing the dimension of an 
arbitrary irreducible representation $\lam$ of $\suN$ 
(with row lengths $\ell_i$ and column lengths $k_i$)
as the determinant of an 
$\ell_1 \times \ell_1$ matrix 
(eq.~(A.6) of ref.~\cite{Fultonbook})
\be
\label{eq:Giambelli}
(\dim \lam)_\suN = \Big| (\dim \Lam_{k_i+j-i})_\suN \Big| \;, 
\qquad i,j = 1, \ldots, \ell_1
\ee
where $\Lam_s$ is the completely 
{\it antisymmetric} representation of $\suN$, 
whose Young tableau is $ \oneoneoneone \} s $.
The maximum value of $s$ appearing in eq.~(\ref{eq:Giambelli})
is $k_1 + \ell_1 - 1$, which is bounded by $N+K-2$
for integrable highest-weight representations of $\suNK$.
The representations $\Lam_0$ and $\Lam_N$ 
both correspond to the identity representation, with dimension 1.
For $1\le s \le N-1$,  
$\Lam_s$ are the fundamental representations of $\suN$,
with $(\dim \Lam_s)_\suN  = \left( N \atop s \right)$.
We define $\dim \Lam_s = 0$ for $s<0$ and for $s>N$.

In ref.~\cite{Naculich:2005tn}, we showed that
\be
\label{eq:twocases}
\!\!\!\!\!\!\!\! (\dim \Lam_s)_{\suN} = 
\left\{  
{
(-1)^s (\dim \tLam_s)_{\suK} \quad \mod \xNK, 
\quad {\rm for~}s \le N+K-2, {\rm~except~} s=N
\atop
(-1)^{K-1} (\dim \tLam_s)_{\suK} \quad \mod \xNK,
\quad {\rm for~}s = N \qquad\qquad\qquad\qquad\qquad
}
\right.
\ee
where $\tLam_s$ is the completely {\it symmetric} representation of $\suK$,
whose Young tableau is $\underbrace{\four}_s$.
(We define $\dim \tLam_s = 0 $ for $s<0$.)
When $N+K$ is odd, eq.~(\ref{eq:twocases}) becomes simply
$(\dim \Lam_s)_{\suN} = (-1)^s (\dim \tLam_s)_{\suK}  \quad \mod \xNK$
for all $s \le N+K-2$.
This was used in ref.~\cite{Naculich:2005tn} 
to yield eq.~(\ref{eq:oddduality}).

Now we turn to the case of even $N+K$,
first considering $N > K$.
In eq.~(\ref{eq:xnk}), the factor $\lcm \{1, \ldots, N-1\}$ 
then contains $(N+K)/2$, so $\xNK$ is at most 2. 
It is easy to see that $\xNK = 2$ if $N+K=2^\power$, and $\xNK=1$ otherwise.
For $\xNK \le 2$, 
the minus signs in eq.~(\ref{eq:twocases}) 
are irrelevant (since $n = -n$ mod 2),
so we may simply write
\be
\label{eq:onecase}
(\dim \Lam_s)_{\suN} = (\dim \tLam_s)_{\suK} \quad \mod \xNK, 
\quad {\rm for~} s \le N+K-2, 
{\rm~with~}N+K{\rm~even~and~}N>K\,.
\ee
We will use this below.

Next we consider $N=K$.
We begin by observing that
if $N$ is a power of a prime $p$,
then $\xNN = 4$ if $p = 2$, and $\xNN = p$ if $p > 2$.
If $N$ contains more than one prime factor, then $\xNN=1$.
In the latter case, eq.~(\ref{eq:evenduality}) is trivially satisfied, 
so we need only consider $N=K=p^\power$, where $p$ is prime.
Let us obtain the relation between 
$(\dim\Lam_s)_{\supm}$ and $(\dim\tLam_s)_{\supm}$ 
by considering three separate cases:
\begin{itemize}
\item $0 \le s \le N-1$:

By examining the factors of $p$ (prime) 
in the numerator and denominator of
$(\dim \Lam_s)_{\supm} = \left( p^\power \atop s \right)$,
one can establish that
if $p^{\subpower-1}$ divides $s$ but $p^\subpower$ does not 
(for any $\subpower \le \power$),
then $p^{\power -\subpower + 1}$ divides $\left( p^\power \atop s \right)$.
Thus $(\dim\Lam_s)_{\supm} = 0$ mod $p$ for $1 \le s \le N-1$.
Combining this with eq.~(\ref{eq:twocases}), we have 
\be
(\dim \Lam_s)_{\supm} = (\dim \tLam_s)_{\supm} \quad \mod \xNN,
\qquad {\rm for~} 1 \le s \le N-1\,.
\ee
This is trivially extended to $s=0$.

\item $s < 0$, or $N+1 \le s \le 2N-2$:

In this case, 
\be
(\dim \Lam_s)_{\supm} = (\dim \tLam_s)_{\supm} \quad \mod \xNN,
\qquad {\rm for~}s < 0, {\rm ~or~}  N+1 \le s \le 2N-2\,,
\ee
is valid because the l.~h.~s.~vanishes, 
and so, by eq.~(\ref{eq:twocases}),
the r.~h.~s.~either vanishes or is a multiple of $\xNN$.

\item $s=N$:

The remaining case yields \cite{Naculich:2005tn}
\be
(\dim \Lam_N)_{\supm} = (-1)^{N-1} (\dim \tLam_N)_{\supm} \quad \mod \xNN
\ee
which is in accord with the other cases when $p$ is a prime other than 2.

\end{itemize}

\noindent We combine these results with eq.~(\ref{eq:onecase})
to write
\bea
\label{eq:allcase}
(\dim \Lam_s)_{\suN} = (\dim \tLam_s)_{\suK} \quad &&\mod \xNK, \quad 
{\rm for~}s \le N+K-2,  \nonumber\\
&&\foreven.
\eea
Inserting this in eq.~(\ref{eq:Giambelli}), we find
\be
(\dim \lam)_\suN 
= \left|  (\dim \tLam_{k_i+j-i})_\suK \right| \quad \mod \xNK,
\quad \foreven.
\ee
By an alternative formula for the dimension of a representation 
(eq.~(A.5) of ref.~\cite{Fultonbook}),
the r.h.s.~is the dimension of a representation of $\suK$ 
with row lengths $k_i$ and column lengths $\ell_i$,\footnote{
If $\lam$ has $\ell_1 = K$,
then the transpose $\tlam$ contains leading columns of $K$ boxes.
In that case, one can apply the formula 
\cite{Maldacena:2001xj}
$Q_{\sig (\lam)} = (-1)^{N-1}  Q_\lam ~\mod \xNK$
several times to relate $\lam$ to a tableau with no rows of length $K$
before using eq.~(\ref{eq:Giambelli}).
The minus sign is irrelevant when $\xNK \le 2$, and vanishes
when $N$ is an odd prime.}
that is, the transpose representation $\tlam$,
hence
\be
\label{eq:dimduality}
(\dim \lam)_\suN = (\dim \tlam)_\suK \quad \mod \xNK,
\quad \foreven.
\ee
from which eq.~(\ref{eq:evenduality}) follows.\footnote{
Since minus signs are irrelevant when $\xNK \le 2$, 
eq.~(\ref{eq:oddduality}) actually holds for all $N\neq K$,
not just odd $N+K$.
Equation (\ref{eq:oddduality}) is not valid, however, when $N=K$.
This is most easily seen by considering representations of $\suNN$
whose tableaux are invariant under transposition, 
and whose dimensions are not multiples of $x$, such
as the adjoint of $\sutt$.}

\vs{.1in}
\noindent{\bf Level-rank duality of the untwisted open string spectrum}
\vs{.1in}

\option
In ref.~\cite{Altschuler:1989nm,Mlawer:1990uv},
it was shown that the fusion coefficients $\Nabc$ of 
the bulk $\suNK$ WZW model are related to those
of the $\suKN$ WZW model, denoted by $\tN$,  by 
\be
\Nabc =  \tNabcD =  {{\tN}_{\tmu \sig^{-\Dr}(\tnu)}}^{~~~~~~~~\tlam } 
\ee
where $\Dr=[r(\mu)+r(\nu)-r(\lam)]/N $.

Since by eq.~(\ref{eq:verlinde}) the fusion coefficients $\Nabc$ are equal 
to the coefficients $\nabc$ of 
the open-string partition function (\ref{eq:openpart}),
it follows that if 
the spectrum of an $\suNK$ open string 
stretched between untwisted D-branes $\lam$ and $\mu$ 
contains $\nabc$ copies of the 
highest-weight representation $V_\nu$ of $\suNK$, then 
the spectrum of an $\suKN$ open string 
stretched between untwisted D-branes $\tlam$ and $\tmu$ 
contains an equal number of copies of the 
highest-weight representation $V_{\sig^{-\Dr}(\tnu)}$ of $\suKN$.

\section{Level-rank duality of untwisted D-branes of $\spnk$}
\renewcommand{\theequation}{4.\arabic{equation}}
\setcounter{equation}{0}

In this section, we examine the relation between untwisted D-branes 
of the $\spnk$ model 
and those of the level-rank-dual $\spkn$ model.

Untwisted D-branes of $\spnk$ are labelled by integrable
highest-weight representations $V_\lam$ of $\spnk = \Conek$.
The D0-brane charge of D-branes of $\spnk$ are defined modulo the 
integer \cite{Bouwknegt:2002bq, Gaberdiel:2003kv}
\bea
x &=&  {  n+k+1 \over \gcd \{ n+k+1, 
\lcm \{1, 2, 3,  \ldots, n, 1, 3, 5, \ldots, 2n-1\} \}  }
 \nonumber \\
  &=&  {  n+k+1 \over \gcd \{ n+k+1, 
{1\over 2} \lcm \{1, 2,  \ldots, 2n\} \}  }
 \nonumber \\
  &=&  {  2(n+k+1) \over \gcd \{ 2(n+k+1), 
           \lcm \{1, 2,  \ldots, 2n\} \}  }
 \nonumber \\
  &=&    \xnk
\eea
where $\xnk$ is given by eq.~(\ref{eq:xnk}).
That is, 
\be
\label{eq:spcharge}
Q_\lam = (\dim \lam)_\spn  \quad \mod   \xnk \qquad {\rm for}~~ \spnk
\ee
is the charge of the untwisted $\spnk$ D-brane labelled by $\lam$,
where $(\dim \lam)_\spn$ is the dimension of the $\spn$ representation $\lam$.
As we showed in the previous section, 
for $n \neq k$,  we have $\xnk =2$ if $n+k+1 = 2^\power$, 
and $\xnk=1$ otherwise.
For $n = k$, we have $\xnn=p$ if $2n+1 = p^\power$, 
and $\xnn=1$ if $2n+1$ contains more than one prime factor.

Since charges of $\spnk$ D-branes are only defined modulo $\xnk$,
and those of $\spkn$ D-branes modulo $\xkn$,
comparison of charges of level-rank-dual D-branes 
is only possible modulo 
$\gcd \{ \xnk,\xkn \} $.
Without loss of generality we henceforth assume that $n \ge k$,
in which case $\gcd \{ \xnk,\xkn \} = \xnk$.

\vs{.1in}
\noindent{\bf Level-rank duality of untwisted D-brane charges}
\vs{.1in}

\option
Given a Young tableau $\lam$ corresponding to an integrable highest-weight
representation of $\spnk$ (with $n$ or fewer rows and $k$ or fewer columns),
its transpose $\tlam$ corresponds to an integrable highest-weight
representation of $\spkn$.
The mapping between representations is one-to-one, 
in contrast to the case of $\suNK$. 

We will show that the relation 
between $Q_\lam$, 
the charge of the $\spnk$ D-brane labelled by $\lam$,
and $\tQ_\tlam$, 
the charge of the level-rank-dual $\spkn$ D-brane labelled by $\tlam$,
is given by 
\be
\label{eq:spnduality}
\tQ_\tlam = Q_\lam  \quad  \mod \xnk\,. 
\ee
The relation (\ref{eq:spnduality})
is nontrivial only
when $\xnk>1$, 
that is, 
when $n\neq k$ with $n+k+1=2^\power$,
or when $n=k$ with $2n+1=p^\power$.

\vs{.1in}
\noindent {\it Proof of eq.~(\ref{eq:spnduality}):} 
We may write the dimension of an 
arbitrary irreducible representation $\lam$ of $\spn$ 
as the determinant of an 
$\ell_1 \times \ell_1$ matrix 
(Prop.~(A.44) of ref.~\cite{Fultonbook}; see also ref.~\cite{Bourdeau:1991uu})
\be
\label{eq:spndet}
(\dim \lam)_\spn  = 
\left| 
\matrix{ 
\chi_{k_1} & 
(\chi_{k_1+1} + \chi_{k_1-1} )
 &\cdots  & 
(\chi_{k_1+\ell_1-1} + \chi_{k_1-\ell_1+1} )
 \cr
\vdots                  &  \vdots & \vdots & \vdots  \cr 
\chi_{k_i-i+1} & 
(\chi_{k_i-i+2} + \chi_{k_i-i} )
 &\cdots  & 
(\chi_{k_1+\ell_1-i} + \chi_{k_1-\ell_1-i+2} )
 \cr
\vdots                  &  \vdots & \vdots & \vdots  \cr 
}
\right| \;, 
\qquad i,j = 1, \ldots, \ell_1
\ee
where $\chi_s = (\dim \Lam_s)_\spn$,
with $\Lam_s$ the completely {\it antisymmetric} representation of 
$\spn$, whose Young tableau is $ \oneoneoneone \} s $.
The maximum value of $s$ appearing in eq.~(\ref{eq:spndet})
is $k_1 + \ell_1 - 1$, which is bounded by $n+k-1$
for integrable highest-weight representations of $\spnk$.
The representation $\Lam_0$ 
corresponds to the identity representation with dimension 1.
For $1 \le s \le n$,  
$\Lam_s$ are the fundamental representations of $\spn$.
(We define $(\dim \Lam_s)_\spn = 0$ for $s<0$ and for $s>n$.)
Also, let $\tLam_s$ be the completely {\it symmetric} 
representation of $\spk$, whose Young tableau is $\underbrace{\four}_s$.
(We define $(\dim \tLam_s)_\spk = 0$ for $s<0$.)

Next, we may use
the branching rules 
$ (\Lam_s)_\suoddn = \oplus_{t=0}^{s} (\Lam_t)_\spn$ (for $s \le n$)
and 
$ (\tLam_s)_\suoddn = \oplus_{t=0}^{s} (\tLam_t)_\spn$ 
of $\suoddn \supset \spn$
to relate 
the dimensions of the fundamental representations of $\spn$ to 
those of the fundamental representations of $\suoddn$:
\bea
(\dim \Lam_s)_{\spn} &=& (\dim \Lam_s)_{\suoddn} 
                      -(\dim \Lam_{s-1})_{\suoddn}  \,, \nonumber\\
(\dim \tLam_s)_{\spk} &=& (\dim \tLam_s)_{\suoddk} 
                      -(\dim \tLam_{s-1})_{\suoddk} \,.
\eea
Using this together with eq.~(\ref{eq:allcase}), we have
\be
(\dim \Lam_s)_{\spn} = (\dim \tLam_s)_{\spk} \quad \mod \xnk,
\qquad {\rm for~}s \le 2n+2k\,.
\ee
We use this in eq.~(\ref{eq:spndet}) to obtain
\be
(\dim \lam)_\spn  = 
\left| 
\matrix{ 
\tchi_{k_1} & 
(\tchi_{k_1+1} + \tchi_{k_1-1} )
 &\cdots  & 
(\tchi_{k_1+\ell_1-1} + \tchi_{k_1-\ell_1+1} )
 \cr
\vdots                  &  \vdots & \vdots & \vdots  \cr 
\tchi_{k_i-i+1} & 
(\tchi_{k_i-i+2} + \tchi_{k_i-i} )
 &\cdots  & 
(\tchi_{k_1+\ell_1-i} + \tchi_{k_1-\ell_1-i+2} )
 \cr
\vdots                  &  \vdots & \vdots & \vdots  \cr 
}
\right| \quad \mod \xnk
\ee
where $\tchi_s = (\dim \tLam_s)_\spk$.
By an alternative formula for the dimension of a representation 
(Prop.~(A.50) of ref.~\cite{Fultonbook}),
the r.h.s.~is the dimension of a representation of $\spk$
with row lengths $k_i$ and column lengths $\ell_i$, 
that is, the transpose representation $\tlam$,
hence
\be
(\dim \lam)_\spn  = (\dim \tlam)_\spk \quad \mod \xnk,
\ee
from which eq.~(\ref{eq:spnduality}) follows.  {\it QED.}

\vs{.1in}
\noindent{\bf Level-rank duality of the untwisted open string spectrum}
\vs{.1in}

\option
In ref.~\cite{Mlawer:1990uv},
it was shown that the fusion coefficients $\Nabc$ of 
the bulk $\spnk$ WZW model are related to those
of the $\spkn$ WZW model by 
\be
\Nabc =  \tNabc\,.
\ee
Since the fusion coefficients $\Nabc$ are equal 
to the coefficients $\nabc$ of the open-string partition function,
it follows that if 
the spectrum of an $\spnk$ open string 
stretched between untwisted D-branes $\lam$ and $\mu$ 
contains $\nabc$ copies of the 
highest-weight representation $V_\nu$ of $\spnk$, then 
the spectrum of an $\spkn$ open string 
stretched between untwisted D-branes $\tlam$ and $\tmu$ 
contains an equal number of copies of the 
highest-weight representation $V_{\tnu}$ of $\spkn$.

\section{Twisted D-branes of WZW models} 
\renewcommand{\theequation}{5.\arabic{equation}}
\setcounter{equation}{0}

In this section we review some aspects of twisted D-branes of the WZW model, 
drawing on 
refs.~\cite{Behrend:1998fd,Fuchs:1999zi,Birke:1999ik, Gaberdiel:2002qa}.
As in section 2, 
these D-branes correspond to possible boundary conditions 
that can imposed on a boundary WZW model.

A boundary condition more general than eq.~(\ref{eq:untwistedconditions})
that still preserves the $\hgK$ symmetry of the boundary WZW model
is 
\be
\label{eq:twconditions}
\left[ J^a(z) - \omega \bJ^a(\bz)\right] \bigg|_{z=\bz} = 0\,,
\ee
where $\omega$ is an automorphism of the Lie algebra $\g$.
The boundary conditions (\ref{eq:twconditions}) 
correspond to coherent states $\twbdy \in \cH^{\rm closed}$ 
of the bulk WZW model 
that satisfy
\be
\label{eq:twmodes}
\left[  J^a_m + \omega \bJ^a_{-m} \right] \twbdy = 0\,, \qquad m\in \Z\,.
\ee
The $\om$-twisted Ishibashi states  $\twishimu$ are solutions 
of eq.~(\ref{eq:twmodes})
that belong to a single sector 
$ V_\mu \otimes \bV_{\om(\mu)^*} $
of the bulk WZW theory,
and whose normalization is given by
\be
\label{eq:twishinorm}
\twishimubra q^H \twishinu =  \delta_{\mu\nu} \chi_\mu (\tau)\,, 
\qquad q = \e^{2\pi i \tau} \,.
\ee
Since we are considering the diagonal closed-string theory
(\ref{eq:diagonal}),
these states only exist when $\mu = \om(\mu)$,
so the $\om$-twisted Ishibashi states are labelled by
$\mu \in \Exp$, where $\Exp \subset \Pplus$ are the integrable
highest-weight representations of $\hgK$ that satisfy $\om(\mu)=\mu$.
Equivalently, $\mu$ corresponds to a highest-weight representation,
which we denote by $\pi(\mu)$, 
of $\gcup$,  
the orbit Lie algebra \cite{Fuchs:1995zr} associated with $\hgK$.

Solutions of eq.~(\ref{eq:twmodes}) that also satisfy the
Cardy conditions are denoted $\om$-twisted Cardy states $\cardya$,
where the labels $\a$ take values in some set $\Bound$.
The $\om$-twisted Cardy states may be expressed 
as linear combinations of $\om$-twisted Ishibashi states
\be
\label{eq:twcardyishi}
\cardya = \sum_{\mu \in \Exp}
{\psi_{\a \pi(\mu)} \over \sqrt{S_{\id\mu}}} \twishimu 
\ee
where $\psi_{\a \pi(\mu)}$ are some as-yet-undetermined coefficients.
The $\om$-twisted D-branes of $\hgK$ correspond to $\cardya$ 
and are therefore also labelled by $\a \in \Bound$.
These states (apparently) correspond \cite{Birke:1999ik}
to integrable highest-weight representations 
of the $\om$-twisted affine Lie algebra $\twhgK$ 
(but see ref.~\cite{Alekseev:2002rj}).

The partition function 
of open strings stretched between $\om$-twisted D-branes $\a$ and $\b$ 
\be
\label{eq:twopenpartition}
\Zop_{\a\b} (\tau) = \sum_{\lam \in \Pplus} \nblama \chi_\lam (\tau) 
\ee
may alternatively be calculated as the closed-string propagator between
$\om$-twisted Cardy states 
\be
\label{eq:twprop}
\Zop_{\a\b} (\tau) = \cardyabra \tq^H \cardyb\,, 
\qquad \tq = \e^{2\pi i (-1/\tau)   }\,.
\ee
Combining eqs.~(\ref{eq:twprop}), 
(\ref{eq:twcardyishi}),
(\ref{eq:twishinorm}),  and
(\ref{eq:modulartrans}),
we find 
\be
\Zop_{\a\b} (\tau) 
= \sum_{\rho \in \Exp}
{ \psi^*_{\a\pirho} \psi_{\b\pirho} \over S_{\id\rho} } \chi_\rho(-1/\tau)
= \sum_{\lam \in \Pplus}
\sum_{\rho \in \Exp}
{ \psi^*_{\a\pirho} S_{\lam\rho} \psi_{\b\pirho} \over S_{\id\rho} }  
\chi_\lam(\tau)\,.
\ee
Hence, the coefficients of the open-string partition function 
(\ref{eq:twopenpartition})
are given by
\be
\label{eq:opencoeff}
\nblama =  \sum_{\rho \in \Exp}
{ \psi^*_{\a\pirho} S_{\lam\rho} \psi_{\b\pirho} \over S_{\id\rho} }  \,.
\ee
Finally, the coefficients $\psi_{\a\pirho}$ 
relating the $\om$-twisted Cardy states
and $\om$-twisted Ishibashi states
may be identified \cite{Birke:1999ik}
with the modular transformation matrices of characters 
of twisted affine Lie algebras \cite{Kacbook},
as may be seen, for example,
by examining the partition function of an open string 
stretched between an $\om$-twisted and an 
untwisted D-brane \cite{Ishikawa:2001zu,Gaberdiel:2002qa}.

\section{Level-rank duality of twisted D-branes of $\suoddnk$} 
\renewcommand{\theequation}{6.\arabic{equation}}
\setcounter{equation}{0}

The finite Lie algebra $\suN$ possesses an order-two automorphism $\omc$
arising from the invariance of its Dynkin diagram under reflection.
This automorphism maps the Dynkin indices 
of an irreducible representation $a_i \to a_{N-i}$, 
and corresponds to charge conjugation of the representation. 
This automorphism lifts to an automorphism of the affine Lie algebra $\suNK$,
leaving the \zeroth node of the extended Dynkin diagram invariant, 
and gives rise to a class of $\omc$-twisted D-branes of 
the $\suNK$ WZW model (for $N>2$).
Since the details of the $\omc$-twisted D-branes
differ significantly between even and odd $N$,
and we will restrict our attention 
to the $\omc$-twisted D-branes of the $\suoddnk = \Annonek$ WZW model.

First, recall that the $\omc$-twisted Ishibashi states $\Ctwishimu$
are labelled by self-conjugate integrable highest-weight 
representations $\mu \in \Exp$ of $\Annonek$.
Equation (\ref{eq:integrable}) implies
that the Dynkin indices 
$(a_0, a_1, a_2, \cdots, a_{n-1}, a_n, a_n, a_{n-1}, \cdots, a_1)$
of $\mu$ satisfy 
\be
\label{eq:twistedintegrable}
a_0 + 2(a_1 + \cdots + a_n) = 2k+1 \,.
\ee
In ref.~\cite{Fuchs:1995zr}, it was shown that 
the self-conjugate highest-weight representations of $\Annonek$ 
are in one-to-one correspondence with 
integrable highest weight representations 
of the associated orbit Lie algebra $\gcup = \Anntwok$,
whose Dynkin diagram is

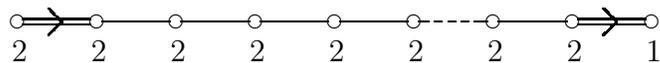
\begin{picture}(500,50)(10,10)
\put(100,40){\circle{5}}
\put(98,25){2}
\put(102,39){\line(1,0){26}}
\drawline(117,40)(112,45)
\drawline(117,40)(112,35)
\put(102,41){\line(1,0){26}}

\put(130,40){\circle{5}}
\put(128,25){2}
\put(132,40){\line(1,0){26}}

\put(160,40){\circle{5}}
\put(158,25){2}
\put(162,40){\line(1,0){26}}

\put(190,40){\circle{5}}
\put(188,25){2}
\put(192,40){\line(1,0){26}}

\put(220,40){\circle{5}}
\put(218,25){2}
\put(222,40){\line(1,0){26}}

\put(250,40){\circle{5}}
\put(248,25){2}
\put(252,40){\line(1,0){4}}
\put(258,40){\line(1,0){3}}
\put(263,40){\line(1,0){3}}
\put(268,40){\line(1,0){3}}
\put(273,40){\line(1,0){5}}

\put(280,40){\circle{5}}
\put(278,25){2}
\put(282,40){\line(1,0){26}}

\put(310,40){\circle{5}}
\put(308,25){2}
\put(312,41){\line(1,0){26}}
\drawline(327,40)(322,45)
\drawline(327,40)(322,35)
\put(312,39){\line(1,0){26}}

\put(340,40){\circle{5}}
\put(338,25){1}

\end{picture}

\option
with the integers indicating the dual Coxeter label $m_i$ of each node.
The representation $\mu \in \Exp$ corresponds
to the $\Anntwok$ representation $\pi(\mu)$ with Dynkin indices
$(a_0, a_1,  \cdots,  a_n)$.
Consistency with eq.~(\ref{eq:twistedintegrable})
requires that the dual Coxeter labels are
$(m_0, m_1, \cdots, m_n)=(1,2,2,\cdots,2)$,
and hence we must choose as the \zeroth node
the {\it right-most} node of the Dynkin diagram above.
The finite part of the orbit Lie algebra $\gcup$,
obtained by omitting the \zeroth node, is thus $C_n$.
(Note that $C_n$ is the orbit Lie algebra of the 
finite Lie algebra $A_{2n}$ \cite{Fuchs:1995zr}.)

Observe that, by eq.~({\ref{eq:twistedintegrable}),
$a_0$ must be odd, 
and that the representation $\pi(\mu)$ of the orbit algebra $\gcup$ 
is in one-to-one 
correspondence \cite{Fuchs:1995zr,Petkova:2002yj,Gaberdiel:2002qa}
with the integrable highest-weight representation $\pi(\mu)'$ 
of the untwisted affine Lie algebra $\Conek$ 
with Dynkin indices $(a_0', a_1', \cdots, a_n')$,
where 
$a_0' = \half (a_0-1)$
and $a_i' = a_i$ for $i=1, \cdots, n$.

Next, the $\omc$-twisted Cardy states  $\Ccardya$
(and therefore the $\omc$-twisted D-branes) 
of the $\Annonek$ WZW model
are (apparently) labelled \cite{Birke:1999ik}
by the integrable highest-weight representations 
$\a \in \Boundomc$ 
of the twisted Lie algebra $\twhgk = \Anntwok$ 
(but see ref.~\cite{Alekseev:2002rj}).
We adopt the same convention as above for the labelling of the nodes 
of the Dynkin diagram 
(consistent with refs.~\cite{Kacbook,Gaberdiel:2002qa}
but differing from refs.~\cite{Goddard:1986bp,Fuchsbuch}).
Thus, the Dynkin indices 
$(a_0, a_1, \cdots, a_n)$ of the highest weights $\a$  
must also satisfy eq.~(\ref{eq:twistedintegrable}),
and the $\omc$-twisted D-branes are therefore characterized 
\cite{Gaberdiel:2002qa,Alekseev:2002rj}
by the irreducible representations of 
$C_n  = \spn$ 
with Dynkin indices $(a_1, \cdots, a_n)$
(also denoted, with a slight abuse of notation, by $\a$).
The charge of the $\om_c$-twisted D-brane  of $\suoddnk$
labelled by $\a$ is given by \cite{Gaberdiel:2003kv}
\be
\label{eq:twcharge}
Q^\omc_\a = (\dim \a)_\spn  \quad \mod \xnk \qquad {\rm for}~~ \suoddnk  \,.
\ee
The periodicity of the charge 
is the same as that of all D-branes of $\suoddnk$.

Observe also that the $\omc$-twisted D-branes $\a \in \Boundomc$ 
are in one-to-one correspondence with
integrable highest-weight representations $\a'$ 
of the untwisted affine Lie algebra $\Conek$ 
with Dynkin indices $(a_0', a_1', \cdots, a_n')$,
where 
$a_0' = \half (a_0-1)$
and $a_i' = a_i$ for $i=1, \cdots, n$.
That is, both the $\omc$-twisted Ishibashi states
and the $\omc$-twisted Cardy states of $\suoddnk$
are classified by integrable representations of $\spnk$.

Recall from eq.~(\ref{eq:opencoeff}) that the coefficients 
of the partition function of open strings 
stretched between $\omc$-twisted D-branes $\a$ and $\b$ 
are given by 
\be
\nblama =  \sum_{\rho \in \Exp}
{ \psi^*_{\a\pirho} S_{\lam\rho} \psi_{\b\pirho} \over S_{\id\rho} } 
\ee
where $\a$, $\b \in \Boundomc$,  $\lam \in \Pplus$,
and $\pirho$ is the representation 
of the orbit Lie algebra $\Anntwok$ 
that corresponds to the self-conjugate representation $\rho$ of $\suoddnk$.
The coefficients $\psi_{\a\pirho}$ are 
given \cite{Birke:1999ik,Ishikawa:2001zu,Gaberdiel:2002qa}
by the modular transformation matrix of the characters of $\Anntwok$.
These in turn may be 
identified \cite{Fuchs:1995zr,Petkova:2002yj,Gaberdiel:2002qa}
with $\hS_{\a' \pirho'}$, the modular transformation matrix of 
$\Conek = \spnk$, so
\be
\label{eq:opencoefftwo}
\nblama =  \sum_{\rho \in \Exp}
{ \hS^*_{\a'\pirho'} S_{\lam\rho} \hS_{\b'\pirho'} \over S_{\id\rho} }  \,.
\ee
We will use this below to demonstrate level-rank duality of $\nblama$.

\vs{.1in}
\noindent{\bf Level-rank duality of twisted D-brane charges}
\vs{.1in}

\option
It is now straightforward to show the equality of charges
of level-rank-dual $\om_c$-twisted D-branes of $\suoddnk$.
As seen above,
the $\om_c$-twisted $\suoddnk$ D-brane labelled by $\a$ 
is in one-to-one correspondence with 
an integrable highest-weight representation $\a'$ of $\spnk$,
and has the same charge (\ref{eq:twcharge}) 
as the untwisted $\spnk$ D-brane labelled by $\a'$ (\ref{eq:spcharge}),
including periodicity.
The integrable highest-weight representation $\a'$ of $\spnk$
is level-rank-dual to the 
integrable highest-weight representation $\ta'$ of $\spkn$
obtained by transposing the Young tableau corresponding to $\a'$,
and the charges of the corresponding untwisted D-branes obey
\be
(\dim \a')_\spn  = (\dim \ta')_\spk \quad \mod \xnk,
\ee
as shown in sec.~4.
Therefore the $\om_c$-twisted D-branes of $\suoddnk$ 
are in one-to-one correspondence with the 
$\om_c$-twisted D-branes of $\suoddkn$,
and the charges of level-rank-dual $\om_c$-twisted D-branes obey 
\be
Q^\omc_\a = \tQ^\omc_\ta \quad \mod \xnk
\ee
where the map between $\omc$-twisted D-branes 
is given by transposition of the associated $\spnk$ tableaux.

\vs{.1in}
\noindent{\bf Level-rank duality of the twisted open string spectrum}
\vs{.1in}

\option
The coefficients of the partition function 
of open strings stretched between $\omc$-twisted D-branes  $\a$ and $\b$
are real numbers so we may write (\ref{eq:opencoefftwo}) as
\be
\nblama =  \sum_{\rho \in \Exp}
{ \hS_{\a'\pirho'} S^*_{\lam\rho} \hS^*_{\b'\pirho'} 
\over S^*_{\id\rho} }  \,.
\ee
Under level-rank duality,
the $\suNK$ modular transformation matrices transform 
as \cite{Altschuler:1989nm,Mlawer:1990uv}
\be 
S_{\lam\mu} 
= \sqrt{K \over N} \e^{-2\pi i r(\lam) r(\mu)/NK} \tS^*_{\tlam\tmu} 
\ee
and the (real) $\spnk$ modular transformation matrices transform 
as \cite{Mlawer:1990uv}
\be
\hS_{\a'\b'} = \htS_{\ta'\tb'} = \htSstar_{\ta'\tb'}
\ee
where $\tS$ and $\htS$ denote the $\suKN$ and $\spkn$ 
modular transformation matrices respectively,
$\tmu$ is the transpose of the Young tableau corresponding 
to the $\suNK$ representation $\mu$,
and $\ta'$ is the transpose of the Young tableau corresponding
to the $\spnk$ representation $\a'$.
These imply
\bea
\label{eq:nblama}
\nblama &=&  \sum_{\rho \in \Exp}
{ \htSstar_{\ta'\widetilde{\pirho'} } \tS_{\tlam\trho} 
\htS_{\tb'\widetilde{\pirho'} } \over \tS_{\id\trho} }  
\e^{ 2\pi i r(\lam) r(\rho)/(2n+1)(2k+1)}  
\nonumber\\
&=&
  \sum_{\rho \in \Exp}
{ \tpsi^*_{\ta\widetilde{\pirho} } \tS_{\tlam\trho} 
\tpsi_{\tb\widetilde{\pirho} } \over \tS_{\id\trho} }  
\e^{ 2\pi i r(\lam) r(\rho)/(2n+1)(2k+1)}   \,.
\eea
Let $\rhohat$ be the self-conjugate $\suoddkn$ representation 
that maps to the $\spkn$ representation $\widetilde{\pirho'}$,
which is the transpose of the $\spnk$ representation $\pirho'$.
In other words, the representation $\pi(\rhohat)$ of the orbit algebra
is identified with $\widetilde{\pirho}$.
Now $\rhohat$ is not equal to $\trho$ (the transpose of $\rho$),
which is generally not a self-conjugate representation,
but they are in the same cominimal equivalence class,
\be
\label{eq:cominimal}
\trho = \sig^{r(\rho)/(2n+1)} (\rhohat) \,,
\ee
which we prove at the end of this section. 
Equation (\ref{eq:cominimal}) implies 
that \cite{Altschuler:1989nm,Mlawer:1990uv} 
\be
\tS_{\tlam\trho} = 
\e^{-2\pi i r(\lam) r(\rho)/(2n+1)(2k+1)}  
\tS_{\tlam\rhohat}
\ee
so that eq.~(\ref{eq:nblama}) becomes
\be
\nblama 
  = \sum_{\rhohat}
{ \tpsi^*_{\ta\pirhohat}  \tS_{\tlam\rhohat} \tpsi_{\tb\pirhohat}  
\over \tS_{\id\rhohat} }  
= \tn_{\tb\tlam}^{~~~\ta}   \,,
\ee 
proving the level-rank duality of the coefficients of 
the open-string partition function of $\omc$-twisted D-branes 
of $\suoddnk$.
That is,  
if the spectrum of an $\suoddnk$ open string 
stretched between $\omc$-untwisted D-branes $\a$ and $\b$ 
contains $\nblama$ copies of the 
highest-weight representation $V_\lam$ of $\suoddnk$, 
then the spectrum of an $\suoddkn$ open string 
stretched between $\omc$-twisted D-branes $\ta$ and $\tb$ 
contains an equal number of copies of the 
highest-weight representation $V_{\tlam}$ of $\suoddkn$.

\vs{.1in}
\noindent {\it Proof of eq.~(\ref{eq:cominimal}):}   
Let $\rho$, a self-conjugate representation of $\suoddnk$,
have Dynkin indices 
\be
\label{eq:rho}
\rho = (2k + 1 - 2 \el, a_1, \cdots, a_n, a_n, \cdots, a_1)
\ee
where $\el = \sum_{i=1}^n a_i$.
The Young tableau for $\rho$ has $r(\rho)= (2n+1)\el$ boxes.
The representation $\pirho'$ of $\spnk$ that corresponds to $\rho$
has Dynkin indices
$(k - \el, a_1, \cdots, a_n)$.
Let the transpose representation 
$\widetilde{\pirho'}$ of $\spkn$
have Dynkin indices
$(n - \tel, \taa_1, \cdots, \taa_k)$,
with $\tel = \sum_{i=1}^k \taa_i$.
The representation $\rhohat$ of $\suoddkn$
that corresponds to  $\widetilde{\pirho'}$
has Dynkin indices
$(2n + 1 - 2 \tel, \taa_1, \cdots, \taa_k, \taa_k, \cdots, \taa_1)$.
Finally, the representation $\sig^\el (\rhohat)$ has Dynkin indices
\be
\label{eq:rhohat}
\sig^\el(\rhohat) 
= (\taa_\el, \taa_{\el-1}, \cdots, \taa_{1}, 2n + 1 - 2 \tel, 
\taa_1, \cdots, \taa_\el, 0, \cdots, 0)
\ee
where the last $2(k-\el)$ entries vanish since $\taa_i=0$ for $i>\el$.

Since $\pirho'$ and $\widetilde{\pirho'}$ are transpose representations,
with row lengths $\ell_i = \sum_{j=i}^{n} a_j$ 
and $\tell_i = \sum_{j=i}^{k} \taa_j$ respectively,
their index sets, defined by \cite{Altschuler:1989nm,Mlawer:1990uv} 
\be 
I = \{\ell_i - i + n+1 ~|~  1 \le i \le n\}, \qquad
\bI = \{n+i-\tell_i    ~|~  1 \le i \le \el \}
\ee
satisfy 
\be
\label{eq:complementary}
I \cup \bI = \{1, 2, \cdots, n+\el \}, \qquad I \cap \bI = 0
\ee
where we have used $\tell_i = 0$ for $i > \el$.

To prove that the Young tableau of $\sig^\el(\rhohat)$ is 
the transpose  of $\rho$, 
we must show that the index sets \cite{Altschuler:1989nm,Mlawer:1990uv}
\be 
J = \{\lam_i - i + 2n + 2 ~|~  1 \le i \le 2n+1\}, \qquad
\bJ = \{2n+ 1 +i-\hlam_i    ~|~  1 \le i \le 2k+1\}
\ee
(where $\lam_i$ and $\hlam_i$ are the row lengths of $\rho$
and $\sig^\el(\rhohat)$ respectively, 
and $\lam_{2n+1}= \hlam_{2k+1}=0$)
satisfy 
\be
\label{eq:quoderat}
J \cup \bJ = \{1, 2, \cdots, 2n+2k+2\}, \qquad J \cap \bJ = 0 \,.
\ee
Using eqs.~(\ref{eq:rho}) and (\ref{eq:rhohat}), 
one gets
\bea
J &=& J_1 \cup J_2 \cup J_3, \hskip1.7in 
\bJ=\bJ_1 \cup \bJ_2 \cup \bJ_3 , \\
J_1 &=& \{   \el +i - \ell_i  ~|~   1 \le i \le n \}  , \hskip0.8in \;
\bJ_1 = \{   \tell_i - i  + \el + 1  ~|~   1 \le i \le \el \} , \nonumber\\
J_2 &=& \{   n + \el   + 1   \} , \hskip1.7in
\bJ_2 = \{   2n + \el + 1 +i - \tell_i  ~|~   1 \le i \le \el \} , \nonumber\\
J_3 &=& \{   2n + 2 + \el + \ell_i - i ~|~  1 \le i \le n \} , \hskip.2in \;
\bJ_3 = \{   2n + 2 \el + 1 + i  ~|~   1 \le i \le 2k-2\el+1 \} , \nonumber
\eea
where $\ell_i$ and $\tell_i$
are the row lengths of the $\spnk$ and $\spkn$ representations
$\pirho'$ and $\widetilde{\pirho'}$.
Using eq.~(\ref{eq:complementary}), one observes that
\bea
J_1 \cup \bJ_1 &=& \{1, 2, \cdots, n+\el \}, 
\hskip1.4in  J_1 \cap \bJ_1 = 0 , \nonumber\\
J_2 &=& \{  n +  \el  + 1   \},  \nonumber\\
J_3 \cup \bJ_2 &=& \{ n + \el + 2, \cdots, 2n+2\el+1 \}, 
\hskip.5in J_3 \cap \bJ_2 = 0 ,\nonumber\\
\bJ_3 &=& \{   2n + 2 \el + 2, \cdots, 2n+ 2k + 2\}  ,
\eea
which establishes eq.~(\ref{eq:quoderat}).  {\it QED.}
\vs{.1in}

\section{Conclusions}
\renewcommand{\theequation}{7.\arabic{equation}}
\setcounter{equation}{0}

In this paper, we have continued our analysis,
begun in ref.~\cite{Naculich:2005tn},
of level-rank duality in boundary WZW models.
We examined the relation between the D0-brane charges of 
level-rank dual untwisted D-branes of $\suNK$ and $\spnk$,
and of level-rank dual $\omc$-twisted D-branes of $\suoddnk$.
We also demonstrated the level-rank duality of the 
spectrum of an open string stretched between 
untwisted or $\omc$-twisted D-branes in each of these theories.
The analysis of level-rank duality of $\omc$-twisted D-branes of $\suoddnk$
is facilitated by their close relation to untwisted D-branes of $\spnk$.

It is expected that level-rank duality will also be present
in the boundary WZW models and D-branes of other level-rank
dual groups.
Also, the level-rank duality of bulk $\suNK$ WZW models
presumably has consequences for the
twisted D-branes of boundary $\suNK$ models 
even when $N$ and $K$ are not odd.
The level-rank map between the twisted D-branes in these cases 
is expected, however, to be more complicated than for $\suoddnk$. 
We leave this to future work.

Further, it would be interesting to derive the level-rank dualities
described in this paper directly from K-theory.

\section*{Acknowledgments}

SGN would like to thank M. Gaberdiel for illuminating comments. 


\vfil\break
\providecommand{\href}[2]{#2}\begingroup\raggedright\endgroup

\end{document}